\documentclass[pr,aps,praps,amsbsy,superscriptaddress]{revtex4}
\usepackage{graphics} 
\usepackage{color}
\begin{document}
\title{Multiple scattering of channeled and
non-channeled positively charged particles in bent monocrystalline silicon.  }

\author{W.~Scandale}
\affiliation{CERN, European Organization for Nuclear Recearch, CH-1211 Geneva 23, Switzerland}
\author{G.~Arduini}
\affiliation{CERN, European Organization for Nuclear Recearch, CH-1211 Geneva 23, Switzerland}
\author{F.~Cerutti}
\affiliation{CERN, European Organization for Nuclear Recearch, CH-1211 Geneva 23, Switzerland}
\author{L.S.~Esposito}
\affiliation{CERN, European Organization for Nuclear Recearch, CH-1211 Geneva 23, Switzerland}
\author{M.~Garattini}
\affiliation{CERN, European Organization for Nuclear Recearch, CH-1211 Geneva 23, Switzerland}
\author{S.~Gilardoni}
\affiliation{CERN, European Organization for Nuclear Recearch, CH-1211 Geneva 23, Switzerland}
\author{R.~Losito}
\affiliation{CERN, European Organization for Nuclear Recearch, CH-1211 Geneva 23, Switzerland}
\author{A.~Masi}
\affiliation{CERN, European Organization for Nuclear Recearch, CH-1211 Geneva 23, Switzerland}
\author{D.~Mirarchi}
\affiliation{CERN, European Organization for Nuclear Recearch, CH-1211 Geneva 23, Switzerland}
\author{S.~Montesano}
\affiliation{CERN, European Organization for Nuclear Recearch, CH-1211 Geneva 23, Switzerland}
\author{S.~Redaelli}
\affiliation{CERN, European Organization for Nuclear Recearch, CH-1211 Geneva 23, Switzerland}
\author{R.~Rossi}
\affiliation{CERN, European Organization for Nuclear Recearch, CH-1211 Geneva 23, Switzerland}
\author{G.~Smirnov}
\affiliation{CERN, European Organization for Nuclear Recearch, CH-1211 Geneva 23, Switzerland}
\author{L.~Burmistrov}
\affiliation{Laboratore de l'Accelerateur Lineaire (LAL), Universite Paris Sud Orsay, Orsay, France}
\author{S.~Dubos}
\affiliation{Laboratore de l'Accelerateur Lineaire (LAL), Universite Paris Sud Orsay, Orsay, France}
\author{V.~Puill}
\affiliation{Laboratore de l'Accelerateur Lineaire (LAL), Universite Paris Sud Orsay, Orsay, France}
\author{A.~Stocchi}
\affiliation{Laboratore de l'Accelerateur Lineaire (LAL), Universite Paris Sud Orsay, Orsay, France}
\author{L.~Bandiera}
\affiliation{INFN Sezione di Ferrara and departamento di Fisica e Scienze della Terra, Universita di Ferrara,  Via Sarogat 1 Blocco C, 44121 Ferrara, Italy}
\author{V.~Guidi}
\affiliation{INFN Sezione di Ferrara and departamento di Fisica e Scienze della Terra, Universita di Ferrara,  Via Sarogat 1 Blocco C, 44121 Ferrara, Italy}
\author{A.~Mazzolari}
\affiliation{INFN Sezione di Ferrara and departamento di Fisica e Scienze della Terra, Universita di Ferrara,  Via Sarogat 1 Blocco C, 44121 Ferrara, Italy}
\author{M.~Romagnoni}
\affiliation{INFN Sezione di Ferrara and departamento di Fisica e Scienze della Terra, Universita di Ferrara,  Via Sarogat 1 Blocco C, 44121 Ferrara, Italy}
\author{F.~Murtas} 
\affiliation{CERN, European Organization for Nuclear Recearch, CH-1211 Geneva 23, Switzerland}
\affiliation{INFN, LNF, Via Fermi, 40 00044 Frascati (Roma) , Italy}
\author{F.~Addesa} 
\affiliation{INFN Sezione di Roma Roma, Plazzale Aldo Moro 2, 00185 Rome, Italy}
\author{G.~Cavoto} 
\affiliation{INFN Sezione di Roma Roma, Plazzale Aldo Moro 2, 00185 Rome, Italy}
\author{F.~Iacoangeli}
\affiliation{INFN Sezione di Roma Roma, Plazzale Aldo Moro 2, 00185 Rome, Italy}
\author{F.~Galluccio}
\affiliation{INFN Sezione di Napoli, Italy}
\author{A.G.~Afonin}
\affiliation{NRC Kurchatov Institute - IHEP, 142281, Protvino, Russia} 
\author{Yu.A.~Chesnokov}
\affiliation{NRC Kurchatov Institute - IHEP, 142281, Protvino, Russia} 
\author{A.A.~Durum}
\affiliation{NRC Kurchatov Institute - IHEP, 142281, Protvino, Russia}
\author{V.A.~Maisheev\footnote{Corresponding author: maisheev@ihep.ru} }
\affiliation{NRC Kurchatov Institute - IHEP, 142281, Protvino, Russia} 
\author{Yu.E.~Sandomirskiy}
\affiliation{NRC Kurchatov Institute - IHEP, 142281, Protvino, Russia}
\author{A.A.~Yanovich}
\affiliation{NRC Kurchatov Institute - IHEP, 142281, Protvino, Russia}
\author{A.D.~Kovalenko}
\affiliation{Joint Institute for Nuclear Research, Joliot-Curie 6,141980, Dubna, Russia}  
\author{A.M.~Taratin}
\affiliation{Joint Institute for Nuclear Research, Joliot-Curie 6,141980, Dubna, Russia}  
\author{A.S.~Denisov}
\affiliation{Petersburg Nuclear Physics Institute named by B.P.Konstantinov of NRC "Kurchatov Institute" 188300, Russia, Leningradskaya Oblast,
Gatchina, 1, mkr. Orlova roshcha }
\author{ Yu.A.~Gavrikov} 
\affiliation{Petersburg Nuclear Physics Institute named by B.P.Konstantinov of NRC "Kurchatov Institute" 188300, Russia, Leningradskaya Oblast,
Gatchina, 1, mkr. Orlova roshcha }
\author{Yu.M.~Ivanov}
\affiliation{Petersburg Nuclear Physics Institute named by B.P.Konstantinov of NRC "Kurchatov Institute" 188300, Russia, Leningradskaya Oblast,
Gatchina, 1, mkr. Orlova roshcha }
\author{L.G.~Malyarenko}
\affiliation{Petersburg Nuclear Physics Institute named by B.P.Konstantinov of NRC "Kurchatov Institute" 188300, Russia, Leningradskaya Oblast,
Gatchina, 1, mkr. Orlova roshcha }
\author{J.~Borg}
\affiliation{ Imperial College, London, United Kingdom}
\author{T.~James}
\affiliation{ Imperial College, London, United Kingdom}
\author{G.~Hall}
\affiliation{ Imperial College, London, United Kingdom}
\author{M.~Pesaresi}
\affiliation{ Imperial College, London, United Kingdom}


\begin{abstract}
We present the results of an experimental study of multiple scattering of positively charged high-energy particles in bent samples of monocrystalline silicon. This work confirms the recently discovered effect of a strong reduction 
in the rms multiple scattering angle of particles channeled in the silicon (111) plane. The effect is observed in the plane orthogonal to the bending plane.
We show in detail the influence of angular constraints on the magnitude of the effect. Comparison of the multiple scattering process at different
energies indicates a violation of the law of inverse proportionality  of the rms angle of channeled particles with energy.
By increasing the statistics, we have improved the results of multiple scattering measurements for particles moving, but not channeled, in silicon crystals. 
\end{abstract}
\maketitle 
\section{Introduction}

It has long been known that the interaction of charged particles
with monocrystalline material differs in many respects from
such interactions with an amorphous substance \cite{TM,HU,BKS}. In
particular, in such crystals, the effect of planar channeling
is observed, when positively charged particles entering the
crystal at small angles with respect to the system of crystallographic
planes are captured by these planes between
adjacent layers of atoms \cite{JL}. In the plane orthogonal to the
crystallographic planes the trajectory of particles is determined
by the atomic planar potential. In the direction along
the crystallographic planes the particle motion is practically
independent of the atomic potential but in principle the particle
may interact with individual positively charged atomic
nuclei. The result of such interactions is multiple scattering
of the particle.

In our recent articles \cite{WS0, TAR}, we reported the observation of
    a strong reduction in multiple scattering when
channeling high-energy positively charged particles in the bent silicon (111) and (110) planes.
Moreover, the effect takes place in the transverse plane of the crystal, perpendicular to the bending plane.
The experiments measured the value of the rms scattering angle of channeled and non-channeled particles in the  direction
  orthogonal to the bending plane of the  crystal.
According to the measurements, the rms multiple scattering angle  of channeled particles was 4-6 times
less than non-channeled particles. It should be noted that the  experiments performed differed noticeably from each other.
In particular, in experiment \cite{WS0}, there was practically no restriction of particles from the input geometric parameters of the particle beam,
whereas in experiment \cite{TAR}, we selected particles with angles of entry into the crystal significantly less than the critical channeling angle.
The comparison of the results of these two experiments stimulated us to further study the effect, in particular to investigate the influence
of the geometric parameters of the beam on the rms scattering angles.

Thus, this new study of ours is based on the experimental data obtained earlier in the experiments with  focusing crystals \cite{WS0}.
A detailed description of the experiment can be found in the articles \cite{WS0,WS}  Here we will give only the main details and motives of this experiment that are important for understanding.

The need for a new, more detailed study is due to the fact that
that the first experiments demonstrated the existence of the effect of reduction of 
multiple scattering but did not show the influence on the effect of various input 
parameters of the particle beam that are captured in the channeling mode. One of the main parameters
are different angular characteristics of the channeling particles in the crystal. 
There is no such information in the description of the experiment \cite{WS0},
and the experiment\cite{TAR}  was performed for a strongly collimated particle beam.  
Thus, obtaining quantitative data on multiple scattering is necessary for a complete understanding 
of this process for insertion into programs for calculating the interaction of particles with single crystals. 

Bent crystals are expected to be used in various high energy physics applications.
In particular, it is planned to use them for collimation and extraction of proton and ion beams at the LHC collider\cite{LHC}.
For this, programs for calculating the passage of particles through crystals have been created, and new results 
can be used in these programs.

 Additionally we would like to point out that the observed  effect has a simple and clear explanation. In amorphous
media relativistic particles move at a range of distances
(impact parameters) relative to atomic centers (nuclei). This
range of impact parameters varies approximately from the
radius of the screening of the electric field of the atom ($\approx 0.2$ angstrom for silicon) to the
radius of the atomic nucleus $R_n \sim  5\cdot 10^{-13}$ cm.
A particle captured in the planar channeling regime moves
in a space between two adjacent crystallographic planes. At
least some of the channeled particles will not be able to get
close to the nuclei and hence the scattering angle will be
less than in amorphous media.

The observed effect can be attributed (classified) to a large number of phenomena accompanying the channeling of particles in crystals.
Such phenomena include Rutherford scattering, energy-loss processes,
 Rutherford scattering, energy-loss processes,
secondary-electron emission, nuclear reactions, x-ray and gamma-ray
production. All of these processes have cross sections
which depend on the impact parameters  involved in
collisions with individual target atoms.
A characteristic common feature of such processes is the suppression of the probability of their manifestation
in comparison with non-channeling particles.
Nevertheless, it is important to note that the quantitative description of such processes is of a specific nature,
i.e. there are no universal relations for their unified mathematical representation. A description of many of these
processes can be found in the literature  \cite{BKS, JL, Gem, Ot, BKC}.

In addition, for crystal 4, measurements of the rms multiple scattering angle demonstrate  unusual behavior
as a function of thickness in the range from 30 to 50 millimeters, namely this value decreases with increasing thickness.
  The article proposes an explanation of this phenomenon as a result of two competing processes of dechanneling and scattering.
In this paper, we will try to provide evidence to explain the observed paradox.

Note that most of the calculated and experimental studies of multiple scattering are related to amorphous substance. However, we found several theoretical articles devoted to such scattering of non-channeled particles in monocrystalline media\cite{BG,LP, SH, TIK}. These works indicate that there are differences between these processes. However, it follows from these works that one should not expect a significant difference. So the work \cite{LP} indicates the suppression of the angle of multiple scattering by about 10 percent. Unfortunately, the ratios presented in the indicated works are difficult to use for comparison with experiment.  The first to describe this effect was  M.Ter-Mikaelyan in his 1953 year work, i.e. before experimental channeling observation \cite{Ter}.
An experimental observation of the effect mentioned here can be found in the work.\cite{Mazo}.

Precise measurement of multiple scattering of non-channeling particles can shed light on the Ter-Mikaelyan effect
and is also of practical importance since many experiments use silicon wafers in detectors.

 It is important to note in this article (as in the previous ones) we represent results of
measurement of the root-mean-square scattering angle in one plane. Naturally, the charged particle is
scattered in two orthogonal planes (perpendicularly and along the channeling plane). However, it is not
possible to measure multiple scattering in the channeling plane in a sufficiently thick crystal, due to the
oscillating motion of the particle. Moreover, such a measurement is also practically impossible in a
straight (unbent) crystal due to the difficulty of separating channeling and non-channeling particles with
angles smaller than the critical one.

The paper is organized as follows. First, we give a short
description of the experiments with focusing  crystals and describe the procedure
of the data analysis. In the subsequent sections we
present the experimental results obtained with three crystals
in the focusing mode, after which a discussion
and conclusions follow.

\section{About experiments with focusing crystals}

The experimental study of focusing  crystals was carried out at the H8 beam line of
the CERN SPS using an almost  pure 400 GeV/c proton
beam and a 180 GeV/c beam of positive secondary
particles for the measurements.
Five pairs
of silicon microstrip detectors, two upstream and three
downstream of the crystal deflector, were used to measure
incoming and outgoing angles of particles with an angular
resolution in each arm of about 3 $\mu$rad \cite{MP, GH}.

Fig. 1 illustrates
the operation principle of such devices. The focusing crystal
is represented by the sum of rectangle ABCF and triangle
FCD (see Figs. 1a and 1c). Positively charged particles entering the
bent crystal in the channeling regime are deflected through
the same angle over the distance BC (AF). For a sufficiently
large deflection angle, the channeled and non-channeled particles
(background) are spatially separated.
The triangular part of the crystal deflects particles with different
transverse coordinates $x$ according to a linear relationship
between the angle and coordinate. Therefore, the particle
trajectories converge at some (focal) point (see Fig. 1b).

The list of   focusing crystals investigated is presented in
Table 1 of \cite{WS}. For the present study we use the data collected
for crystals 1, 3 and 4. Crystal 1 has dimensions:
$AB = 2.07 \pm 0.01$ mm, $AD = 49.84 \pm  0.02$ mm  $BC =
29.8$ mm. Crystal 3 is approximately the same size. Crystal 4
has the same AD and BC sizes as crystal 1 but AB = 4 mm.
The height (not shown in the figure) of every crystal was 86 mm.

For particles  channeled  in (111) silicon planes, the critical angle was equal to 10.6 $\mu$rad
and 15.8 $\mu$rad for 400 GeV/c and 180 GeV/c beams, respectively. The bending radii are
approximately equal
to 60, 45, 200 m for crystals 1, 3, 4, respectively.

As a result, for
every particle crossing the oriented crystal we obtained: (a)
the horizontal and vertical coordinates; (b) the horizontal and
vertical incident angles; (c) the horizontal and vertical outgoing
angles after the crystal.
The difference between the
horizontal (vertical) outgoing and incoming angles gives the
horizontal (vertical) deflection angle for each particle. Fig.
1b illustrates the results in a two-dimensional plot of the
horizontal angle of deflection versus the horizontal coordinate.
The particles captured in the channeling regime and
which passed through the body of crystal in this regime are
located between lines $Q_+ Q'_+$
 and $Q_- Q'_-$.
 Selected in this
way, the set of channeled particles undergoes multiple scattering
in the vertical direction. Distributions of channeled and non-channeled
particles over vertical scattered angles constitute one of the
subjects of our study.

In addition, the data obtained for each case (channeled
and non-channeled particles) were divided into 21 parts for
crystals 1 and 3 (41 parts for the crystal 4) according to their
horizontal coordinates. So, in section 0 we took particles with
horizontal coordinates from $- 0.05$ to $0.05$ mm, in section 1
those with horizontal coordinates from 0.05 to 0.15 mm, in
section $-1$ those with horizontal coordinates from $-0.05$ to
$- 0.15$ mm and so on. Such a selection of the data allowed
us to study the process of multiple scattering for different
thicknesses.

The linear connection
between $x$ and $z$-coordinates (see Fig. 1c) was
\begin{equation}
z [\mathrm{mm}] = 40 + x_0 [\mathrm{mm}] + kx [\mathrm{mm}]
\end{equation}
where the coefficient $k$ is equal to 9.7 for crystals 1 and 3,
and 5 for crystal 4. The variable $x$ was from -1 to 1 mm
for crystals 1 and 2, and from -2 to 2 mm for crystal 4
($x_0 \approx 0$).

It should be noted that our collaboration (UA9) has carried
out measurements of multiple scattering of 400 GeV/c protons
\cite{WS4} in single silicon crystals with orientations far from
axial or planar channeling. That experiment was performed
in parallel and at the same time and on the same installation as
the experiments described here with focusing crystals. In \cite{WS4}
the background conditions were investigated. They showed
that there is additional scattering of the protons on material
in the beam (strip detectors and other matter). Background
measurements in the experiment performed without a crystal
show that the contribution of this background process can
be described by a Gaussian with rms $\sigma_{bg} = 5~\mu$rad for a
400 GeV/c proton \cite{WS4} and $\sigma_{bg} = 11.27~\mu$rad \cite{TAR} for 180 GeV/c secondary  beams.
As explained in \cite{WS0,TAR, WS4} the measured $\sigma_{m}$  and corrected $\sigma_{si}$ rms
of angles of multiple scattering are determined by the relation:
\begin{equation}
\sigma_{si}^{2} \approx \sigma_{m}^2-\sigma_{bg}^2
\end{equation}
where $ \sigma_{m} $ is the rms measurement resolution of the particle planar angle  with the help of the silicon strip detectors.
For definiteness, we will consider the case when the bending of the crystal lies in the horizontal plane $xz$
($x$ and $z$, respectively, the transverse and longitudinal coordinates of the particle, see Fig. 1). Basically, we will study
scattering in the plane orthogonal to the $xz$ plane (i.e. in the vertical plane).

\section{Analysis}
Multiple scattering of charged particles in an amorphous homogeneous medium has been considered in many articles and monographs (see, for example \cite{TM,BKS,PDG}). It was shown that in such a medium the distribution function of particles over small planar angles is described by the Gaussian distribution
\begin{equation}
\rho(\theta)= {dN\over d\theta} = {1\over \sqrt{2\pi} \sigma}  \exp{-{(\theta -\bar{\theta})^2\over 2\sigma^2 } },
\end{equation}
where $\theta$ is the deflection angle relative to the mean angle $\bar{\theta}$. The value $\sigma$ is the rms of the distribution.

There are different formulas for the rms angle.
 For this study we apply the recommendation
of Ref. \cite{PDG} to use a Gaussian approximation
for the central 98\% of the projected angular distribution, with
the rms equal to:
\begin{equation}
\sigma_0= {13.6 [\mathrm{MeV}] \over \beta c p} \sqrt{l/X_0}[1+0.038\ln(l/X_0)]
\end{equation}
where $p$ and $\beta c$ are the momentum and velocity of the incident particle, $c$ is the velocity of light, $l, X_0$ are the thickness of the scattering medium and its radiation length.

Hence it can be seen that for a parallel beam of relativistic particles of fixed energy moving in an amorphous medium
the process of multiple scattering depends only on the properties of this medium and does not depend on any initial conditions for the particle.
However, this is not valid for the process of multiple scattering (in the vertical plane) of channeled particles.
For example, channeled particles can have different oscillation amplitudes and therefore approach to within different distances
from the atomic centers (nuclei) in the plane. Another example is the dependence on the initial angle of entry of particles into the crystal and departure from it.

 In a real experiment, the scattering of a channeled particle can be influenced by many factors. We  show the effect with a simplified example. Let the rms scattering angle depend on one parameter $\tau$
(for example, the angle of the initial entry of the particle into the crystal
in the horizontal plane). Then, in the general case, for a particle with this parameter, there is an rms scattering angle,
which can be considered
as a function  $\sigma(\tau)$ of this parameter. In addition, the function $h(\tau)$ describing the particle density distribution as a function of $\tau$ is naturally introduced.
Then we  get a distribution function over $\theta$ which takes into account the influence of $\tau$.
\begin{equation}
{\rho (\theta)} ={1\over \sqrt{2\pi} } \int_{\tau_{min}}^{\tau_{max}} {1\over \sigma(\tau)} \exp[{-\theta^2/(2\sigma^2(\tau))}] h(\tau) d\tau
\end{equation}
where  $\tau$ varies from $\tau_{min}$  to $\tau_{max}$.
Under the assumption of a small change in the rms scattering angle $\sigma(\tau) $ , we can write a relation
$\sigma(\tau) = \sigma(0) + \nu(\tau)$     where the condition  $  \nu(\tau) /\sigma(0) \ll  1 $ for the function  $  \nu(\tau)$ is satisfied.  Taking this into account  we obtain for the distribution function
\begin{eqnarray}
{\rho (\theta)}\approx 1/(\sqrt{2\pi}\sigma(0))\exp[{-\theta^2/(2\sigma^2(0))}]
\int_{\tau_{min}}^{\tau_{max}}
(1+\nu(\tau)/ \sigma(0) )\exp[\theta^2\nu(\tau)/ \sigma^3(0)]h(\tau) d\tau
\end{eqnarray}
Taking the first terms of the exponential expansion under the integral we obtain a further simplification
\begin{eqnarray}
{\rho (\theta)}\approx1/(\sqrt{2\pi}\sigma(0))\exp[{-\theta^2/(2\sigma^2(0))}]
\int_{\tau_{min}}^{\tau_{max}}
(1+\nu(\tau)/\sigma(0) + \theta^2\nu(\tau)/\sigma^3(0) ) h(\tau) d\tau
\end{eqnarray}
It is easy to see that for the specified smallness of $\nu$  the integral in Eq.(7)\textsc{} is approximately equal to 1 (with unit normalization of the function  $h(\tau)$).
Hence it follows that under the above assumptions, the shape of the particle multiple scattering distribution remains approximately Gaussian. 
Our new analysis of the data collected in the experiment will be based on a study of the influence of various geometric parameters of the beam on the value of the rms scattering angle. And the content of the  equations obtained justifies the Gaussian approximation of the beam particle distribution after scattering.

Note that for an amorphous medium  $h(\tau)= \delta(0)$ and the Gaussian function for small angle multiple scattering follows from Eq. (7).



 It is also necessary to point out that in a real experiment at sufficiently large scattering angles,
individual hard scattering processes begin to dominate.
Therefore, the process of small-angle multiple scattering should be separated from such processes.
In article \cite{PDG}
for this purpose, it is proposed to take into account only 98\% of the particle angular distribution.
 We accept this general recommendation, and for more concreteness we consider the distribution of particles
in the range from -2.5 to 2.5 standard deviations relative
to the average scattering angle. As a rule, this corresponded to 96-98\%  of the angular distribution.

\section{Multiple scattering  of non-channeled particles}
In our previous paper \cite{WS0} we presented the results of measurements of the vertical angles of multiple scattered particles  at crystal thicknesses from 30 mm to 50 mm.
In this work, we present a) the same results but with higher statistics, which improved the measurement accuracy
significantly, and b) for the first time present the results of measurements of multiple scattering of
non-channeled particles in the horizontal plane (bending plane).

To find the rms angle of non-channeled particles, we used their vertical angular distributions. Particles were selected with an angle of entry into the crystal that
exceeded the critical channeling angle by at least a factor 2. We used data only for
particles with an entry angle  in the opposite direction
to the deflection angle  due to bending.

Fig. 2 illustrates the results of measurements of the vertical rms multiple scattering angle  at
400 GeV/c (crystal 1)  and 180 GeV/c  (crystals 3 and 4).
For comparison the data for  the first crystal were multiplied  by a factor 400/180 in accordance with  Eq.(4). 

We also measured multiple scattering in the horizontal plane for the above-barrier particles (see Fig. 3).
For this purpose, we selected particles with angles from 20 to 45 $\mu$rad relative to the optimal entrance direction for channeling.

As for the vertical angles we used particles with entry angles in the direction opposite to the deflection angle due to  bending.

The results of measurements were approximated by linear functions of the crystal thickness.
It should be noted that
according to theory and, in particular, our measurements, the mean square of the angle of multiple scattering of particles
in an amorphous medium  is a linear function of thickness to good accuracy. However, for the rms scattering angle  in a region of sufficiently large thickness, a linear function is also a good approximation.

\section{Multiple scattering  in the channeling region}

In our previous work, when measuring the scattering of channeled particles, we did not try to strongly limit
the geometric parameters of the beam. It seemed to us that this could create a false reduction in the observed effect.
 In those measurements, we demonstrated for the first time a decrease in the multiple scattering angle of channeled particles
as compared to non-channeled ones. In the experiment, we recorded a decrease in the mean-square multiple scattering angle  of
 channeled particles by  a factor of about 3.

After  publication we continued to study the data collected in the experiment and found that
by changing the beam parameters it is possible to change the value of the rms scattering angle.
Especially noticeable was the influence of the angle between the lines $Q_+\, Q '_+$ and $Q_ -\, Q'_ -$.
 As can be seen from Fig. 1, the gap between the lines $Q_+\, Q '_+$ and $Q_ -\, Q'_ - $ contains particles
that have passed the entire thickness of the crystal in
the channeling regime. We have divided this area into several narrow strips with sides parallel to the straight line $ Q \, Q' $.
This straight line corresponds to the center of one of the strips, which we take as the zero level. The width of each strip was
10 $\mu$rad. In the lower direction relative to the zero level,
the angular magnitude at the center of each strip increases, and vice versa in the opposite direction.
For each strip, taking into account only  particles inside it, we found the rms scattering angle as a function of the crystal thickness.
These results are presented in Figs. 4 and 5 for 400 GeV/c and 180 GeV/c, respectively.
In the interval 30 mm $< z < 50$ mm, the functions $\sigma_c$
(see Figs. 4 and 5) were approximated by the linear function
$const + A_0 z$ for each strip.

Fig. 6 shows the distribution of channeled particles depending on the angle relative to the straight line $ Q \, Q '$.
The part of the distribution  on the left shows the tail of  dechanneled particles. The distribution for 180 GeV/c  differs from the distribution for 400 GeV/c
due to a) a larger bending radius of the crystal and b) a smaller (by more than a factor 2)  dechanneling length.

For 400 GeV/c particles, the rms angle $ \sigma_c$ increases with increasing length for all but one of the strips, which is at an extremity.
This fact corresponds to values of the coefficient $A_0 > 0$.
The results for the central strips (-10, 0, 10) are similar to each other.

For 180 GeV/c momentum, the behavior of the dependence of the scattering angle  differs from the previous case. 
These differences are that
a)  only for particles in strips (-5, 0, 5) is the value of the scattering angle an increasing function of the thickness, and
b) for the extreme strips on both sides, this value is a decreasing function of the thickness.

Note that for some measurement results the linear function is not an adequate approximation, but
we have shown these approximations in the figure because they correctly indicate the trend of the of these results.

Figs. 7 and 8 show rms multiple scattering angles as a function of crystal thickness. Here are the results
for particles falling into the range of angles determined by the condition $ - \phi_o <\phi < \phi_o $.
It can be seen that for particles with a momentum of 400 GeV/c and for the same thickness, the rms angle changes little with $\phi_o$.
In addition, this parameter undergoes a slight bend at coordinates in the vicinity of 40 mm and a weakly pronounced
maximum at a coordinate of about 45 mm.
On the whole, all the results obtained here indicate a steady increase in the scattering angle with increasing thickness.

For 180 GeV/c particles, the situation is somewhat different. The scattering angle of particles with an angle $\phi_o$ (relative to the zero level) less than 16 $\mu$rad
increases slowly with increasing crystal thickness, while at large angles a decrease in the rms scattering angle with thickness is observed.
The coefficient $A_0= 0.053\pm 0.015, 0.038 \pm 0.01, 0. \pm 0.01, -0.065 \pm 0.014 \,\mu$rad/mm
 for curves $\pm 8,\pm 16, \pm 32 ,\pm 48\,\mu$rad, respectively.

We also investigated the value of the rms angle for a narrow beam of particles entering the crystal.
 This beam was defined by the central angle of entry of the particles and an angular width of 5 and 10 $\mu$rad for 400 GeV/c and 180 GeV/c, respectively. These results are shown in Figs. 9 and 10.

Note that in the data presented in this article there are no results corresponding to a coordinate $ x $ near the crystal surface in order to avoid possible surface effects on the measurement results.

\section{Discussion}
\subsection{Multiple scattering of non-channeled particles in the vertical plane}
The  new results we presented from measuring the rms multiple scattering angle of positively charged particles moving
outside the channeling region do not contradict our previous data and are more accurate.
Recall that we are talking about the scattering angle in the transverse plane perpendicular to the crystal bending plane
(i.e. in the vertical plane in the geometry of the experiment).

In our previous study, we showed that the value of the rms angle of multiple scattering of non-channeled particles is slightly less than follows from Eq.(4). However, in order to state this confidently, it was necessary to improve the measurement accuracy. The  new data  confirm this statement.

It follows from  theory that the rms multiple scattering  angle of an ultrarelativistic particle
in an amorphous medium is inversely proportional to its momentum (energy). Our measurements performed for two
different energies confirm this result for non-channeled particles in a silicon monocrystal.
The mean square multiple scattering angle can be considered a linear function of thickness over a relatively small distance of the substance.
These conditions make it possible to write a universal equation for the rms multiple scattering angle in monocrystalline silicon
\begin{equation}
\langle \theta_{si}^2\rangle  = ({\kappa \over E})^2 ( l/X_0+ \delta)
\end{equation}
where coefficients  $\kappa=13.61 \pm 0.13, 12.98 \pm 0.13, 13.04 \pm 0.14 \,\, $MeV/c    and
 $\delta= -0.058 \pm 0.0015, -0.027 \pm 0.0018, -0.045 \pm 0.0019 $ are found from measurements with
crystals 1, 3 and 4, respectively.

Since the condition $ l/ X_0 \gg \delta $ is satisfied in the entire measurement region, we can find the proportionality coefficients
(i.e. neglecting the value of $ \delta $) In this case, we get the results
$\tilde{\kappa} = 12.65\pm 0.29 , 12.57 \pm 0.19, 12.36 \pm 0.18 \,\, $ MeV.
These results are close to our previous measurements, but their accuracy is 1.5-2 times better.

Note that the first measurements of multiple scattering of 400 GeV/c protons were carried out by our collaboration in 2016 \cite{WS4}
for three silicon crystals 0.97, 1.94, and 4.02 mm thick. In the experiment, the crystals were oriented away from strong axes and planes.
We combined the measurement data \cite{WS4} with the data obtained in the experiment \cite{WS0,WS} and obtained a description for the process from fractions of a mm to 50 mm of silicon thickness. For the universal Eq.(8), we got $\kappa = 12.76 \pm 0.09\, $ MeV, $\, \delta=-0.0074 \pm 0.0021$.

Fig. 11 presents various approximations of the mean square scattering angle. It can be seen that a linear relationship describes
experimental points well. The calculation according to Eq.(4) does not agree very well with the measurements.
However, Eq. (4) can be written in the form
\begin{equation}
\sigma_n^2 = [ {\varepsilon  \over \beta c p} ]^2 {z \over X_0}[1+\omega\ln(l/X_0)]^2
\end{equation}
where $\varepsilon$ and $\omega$ are free parameters for the approximation.
 The best fit is for $\varepsilon =13.35$ MeV and $\omega= 0.063$.


\subsection{Multiple scattering of non-channeled particles in the horizontal plane}

For this paper, we studied the scattering behavior in the vertical plane (the plane perpendicular to the bending plane).
However, to complete the picture, we present measurements of the scattering of non-channeled positively charged particles (see Fig. 3)
in the horizontal plane (bending plane).
In contrast to the vertical plane, in this case even non-channeled particles move interacting with the potential of the plane.
As previously indicated, the particles with the angles of entry into the crystal in the direction opposite to the deflection of the channeled
beam were selected for the measurement. This corresponds to the process of planar volume reflection of particles in crystals \cite{TV,MV},
and we observed such a reflection in this experiment.

In the case of volume reflection, the particle trajectory deviates slightly and one can expect that multiple scattering will be
the same as in the vertical plane. We can assume that the angle of deflection of the particle after the crystal will be
is equal to $ \Delta \theta_n = \Delta \theta_m + \Delta \theta_p $, where $ \Delta \theta_m $ is the angle of multiple scattering
at the crystal thickness
and $ \Delta \theta_p $ is the angle of deflection  of the particle due to its interaction with the potential of the plane.
Assuming the independence of these two factors
it is easy to see that the total mean square of the scattering angle in this case should be greater than vertically.
This is demonstrated in Fig. 3.

The experimental points shown in Fig. 3 were approximated by straight lines. It can be seen that the scatter of the experimental
points relative to these straight lines is noticeably larger than the statistical error. One can even see periodic deviations from
straight lines. This suggests the presence of oscillations similar to those observed in  \cite{Sy}.
The problem touched upon here requires further study.

\subsection{Multiple scattering  in the channeling range}
Figs. 4 and 7, and 5 and 8, show changes in the rms multiple scattering angle
for  particles with different angular restrictions relative to the straight line $ QQ '$ (see Fig. 1). Figs. 4 and 5 show the behavior of
the scattering angle in narrow strips, while Figs. 7 and 8 show the behavior of such an angle at different and sufficiently large
particle capture angles. Note that only for crystal 1 and 4 did we have fairly large statistics for particles
in the channeling range. The statistics accumulated on crystal 3 were sufficient for measurements of non-channeled particles
but  poor for particles in the channeling zone.

Particles found in the zone between the straight lines $Q_+\, Q '_+$ and $Q_ -\, Q'_ -$ can be mainly of two types: a) particles that have passed through the entire crystal
in the channeling regime and b) dechanneled particles.

Fig. 6 gives a qualitative representation of the occurrence of dechanneled particles in crystals 1 and 4,
i.e. for 400 and 180 GeV/c particles. In the figure, the zero angular coordinate corresponds to the maximum density of channeled particles.
In the region of large negative angles, we see a tail of dechanneled particles. For crystal 4, this tail has a higher intensity
compared to crystal 1.
This can be explained by the following arguments:
a) the dechanneling length for  crystal 1 is 400/ 180 times greater than for crystal 4,
and therefore, in the first case, fewer dechanneled particles are formed;
b) the particles that dechanneled in the depth of the crystal quickly deviate from the direction of motion of the channeled particles.
It is obvious that for a crystal with a smaller bending radius,  dechanneled particles deviate faster from that direction.

A comparison of Figs. 4 and 5 shows that the rms scattering angle for the strips in the channeling region and away
from the dechaneling region
increases with increasing crystal thickness. This effect is more pronounced for crystal 1, while for crystal 4 it is
much weaker. In the region where dechanneled particles are expected, a decrease in the rms scattering angle is observed
with increasing thickness.
For crystal 1, this effect is observed only for one strip (with a center at -30 $\mu$rads).
Our results show that the process under consideration can be affected by dechanneled particles at certain
crystal parameters and particle energy.

In general, the data presented in Figs. 4 and 5, as well as Figs. 7 and 8, show a reduction  of the rms multiple
scattering angle by 6-8 times in the channeling region compared to non-channeled particles.

Figs. 9 and 10 demonstrate the fact that, at initial entry with angles smaller than the critical channeling angle,
approximate symmetry is observed as a function of the mean square scattering angle. However, for a 400 GeV/c beam
at an angle of incidence of -15 $\mu$rad, this symmetry is broken. This can be explained by the asymmetry of the effective planar  potential.
The absence of such an effect for 180 GeV/c particles can be explained by the large bending radius and, accordingly,
by the smallness of the corresponding addition to the potential.

\subsection{Violation of the energy dependence of multiple scattering}
 As already stated,  theory predicts inverse  proportionality of the rms angle of small-angle multiple scattering with particle energy.
To check this, similarly to our previous article, we plotted the dependence of the square of the multiple scattering angle
divided by $k^2$ ( $k=13.6$[MeV]$/ E]$) as a function of thickness (see Fig. 12). We see that for non-channeled particles these functions coincide with each other
 with good accuracy. For channeled particles, we took the distribution around the zero level to plot. We can see the difference
between these functions in Fig. 12.
The dependence for crystal 3 is also shown here. For this crystal, we took all particles in the range of angles from -40 to 40 $\mu$rad
(due to poor statistics). We see a noticeable difference between the above curves obtained for particles in the channeling region.

We believe that the reason for this violation is that the intensity of the interaction of non-channeled particles with the nuclei in the crystal is
determined by the radiation length which does not depend on the particle energy.
It can be assumed that for channeled particles, the intensity of the interaction is determined by the dechanneling length.
And this value depends approximately linearly on the energy of the particle.

\section{Conclusions}
In this work, we have increased the statistics and also implemented tight control of the beam parameters and introduced restrictions on
angular characteristics of particles that allowed to:

a) confirm the effect of a decrease in the rms scattering angle in comparison with non-channeled particles;

b) obtain a detailed picture of the change in scattering angle with different angular constraints;

c) show that the coefficient of scattering angle reduction is numerically increased up to 6-8 times (in comparison with non-channeled particles);

d) confirm the effect of the unusual behavior of the scattering angle as a function of the thickness. However,
 this is observed  at sufficiently large angles relative to the zero level.
 Although there is not full understanding of this issue, it can be assumed that this behavior is a result of dechanneling processes;

e) find that the behavior of the rms scattering angle does not satisfy inverse proportionality variation with energy;

f) observe some features of multiple scattering of non-channeled particles in the bending plane, which may have an oscillating character; 

g) refine the results of multiple scattering of non-channeled particles;

h) in general, conclude that
our research expands knowledge about channeling and quasi-channeling of high-energy particles in  monocrystalline silicon.
Nevertheless we believe that it is consistent with traditional theoretical views on these processes.

The effect considered here is also of great practical importance. For example, in \cite{CM} an application using two focusing crystals for focusing beams in two orthogonal planes was proposed. It can be expected that the effect of reducing multiple scattering will improve the performance of such a system of lenses. It is also important to understand the scattering of channeled particles in crystals for suggestions to use crystals for laser-plasma acceleration \cite{ST}.

\section{Acknowledgments}
The Imperial College group thanks the UK Science and Technology
Facilities Council for financial support. The INFN authors acknowledge
the support of the ERC Ideas Consolidator Grant No. 615089
CRYSBEAM. The IHEP participants of UA9 experiment acknowledge
financial support of Russian Science Foundation (grant 22-12-00006).
 The corresponding author would  like to thank V.V.  Tikhomirov for
useful discussion.

\begin{figure*} 
\begin{center}
\scalebox{0.8}
{\includegraphics{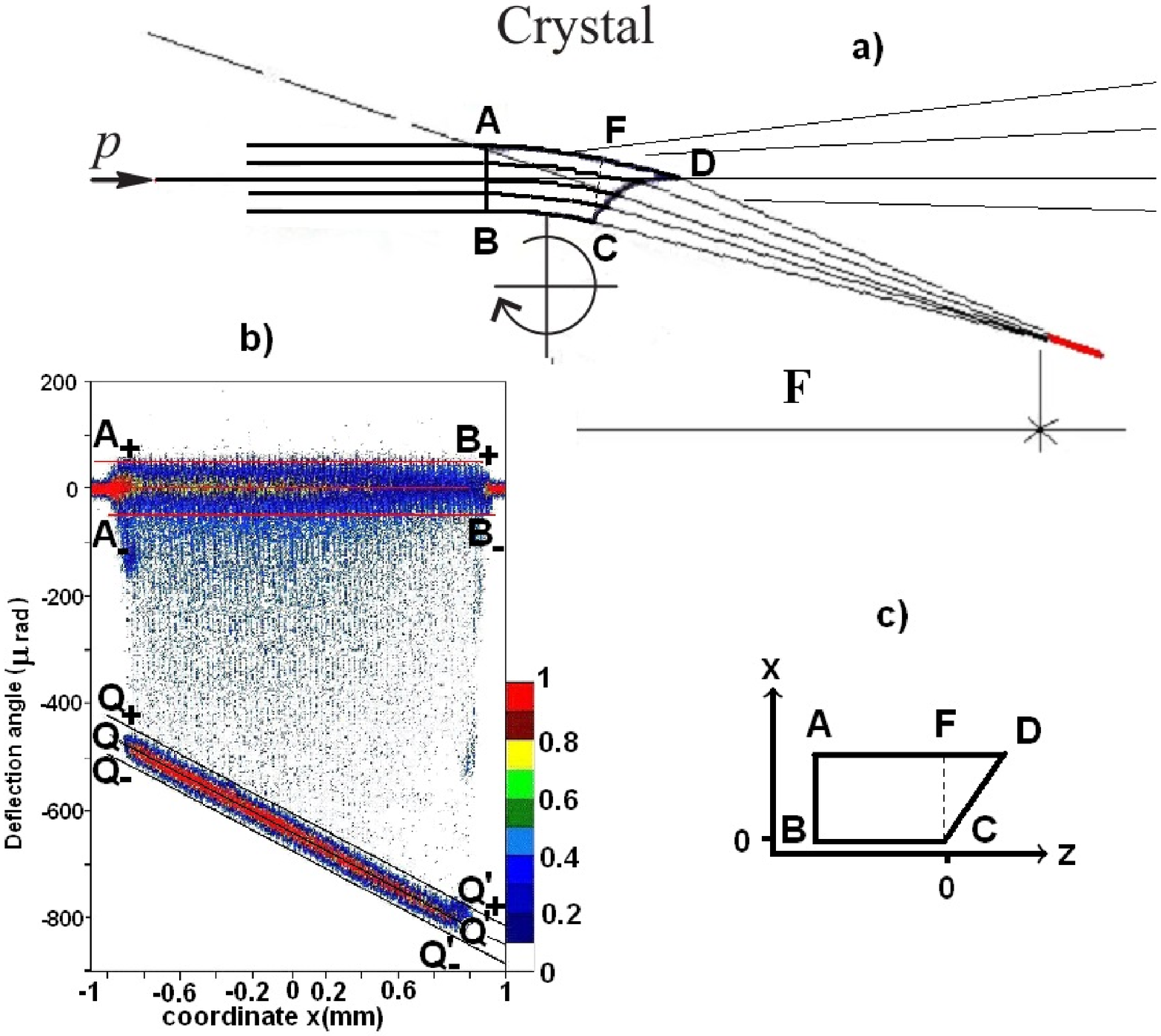}} 
{\caption{
Focusing crystal: a)
operation principle; b) the
measured two-dimensional plot:
deflection angle versus
horizontal coordinate $x$; c)
focusing crystal before
installation in the holder
}}
\end{center}
\end{figure*}

\begin{figure*} 
\begin{center}
\scalebox{0.8}
{\includegraphics{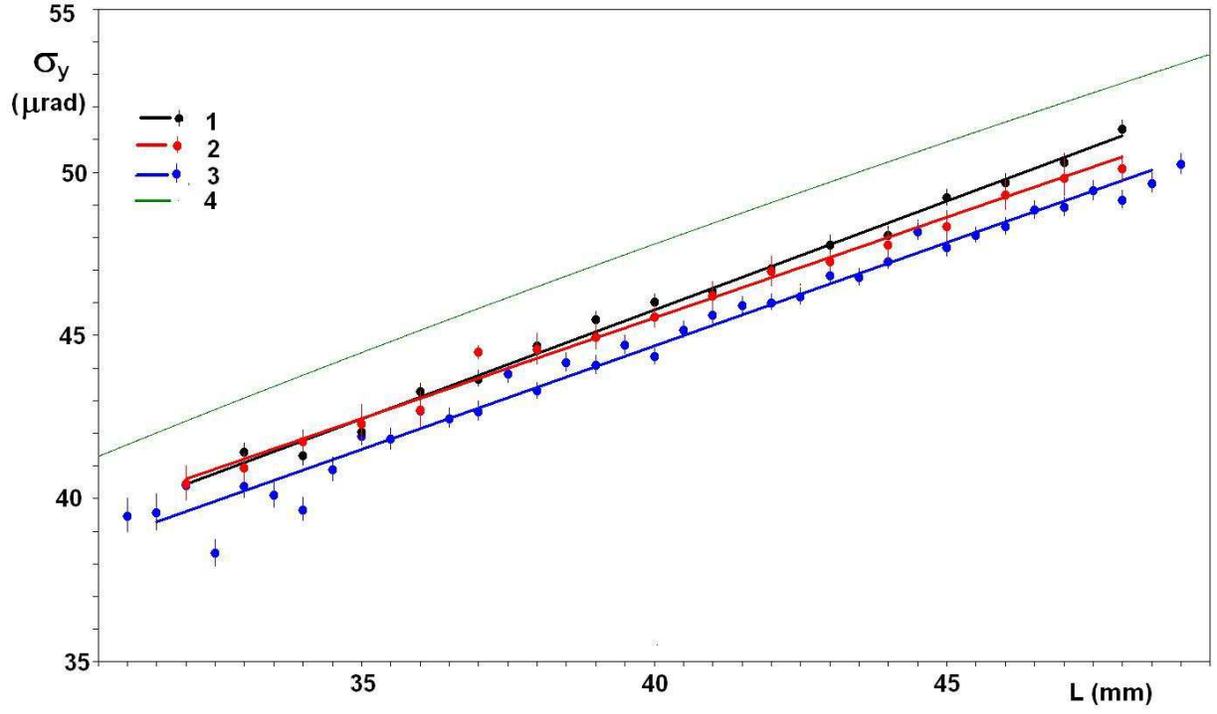}} 
{\caption{
The measured value of the rms angle of positively charged particles moving outside the channeling region.
The curve 1 corresponds to particles with a momentum of 400 GeV/c (crystal 1). For this case, for the convenience of comparison, the measurement results
were multiplied by a factor of 400/180. The curves 2 and 3 correspond to a momentum of 180 GeV/c and crystals 3 and 4, respectively. The curve 4 is a calculation according to Eq.(4).
}}
\end{center}
\end{figure*}

\begin{figure*} 
\begin{center}
\scalebox{0.8}
{\includegraphics{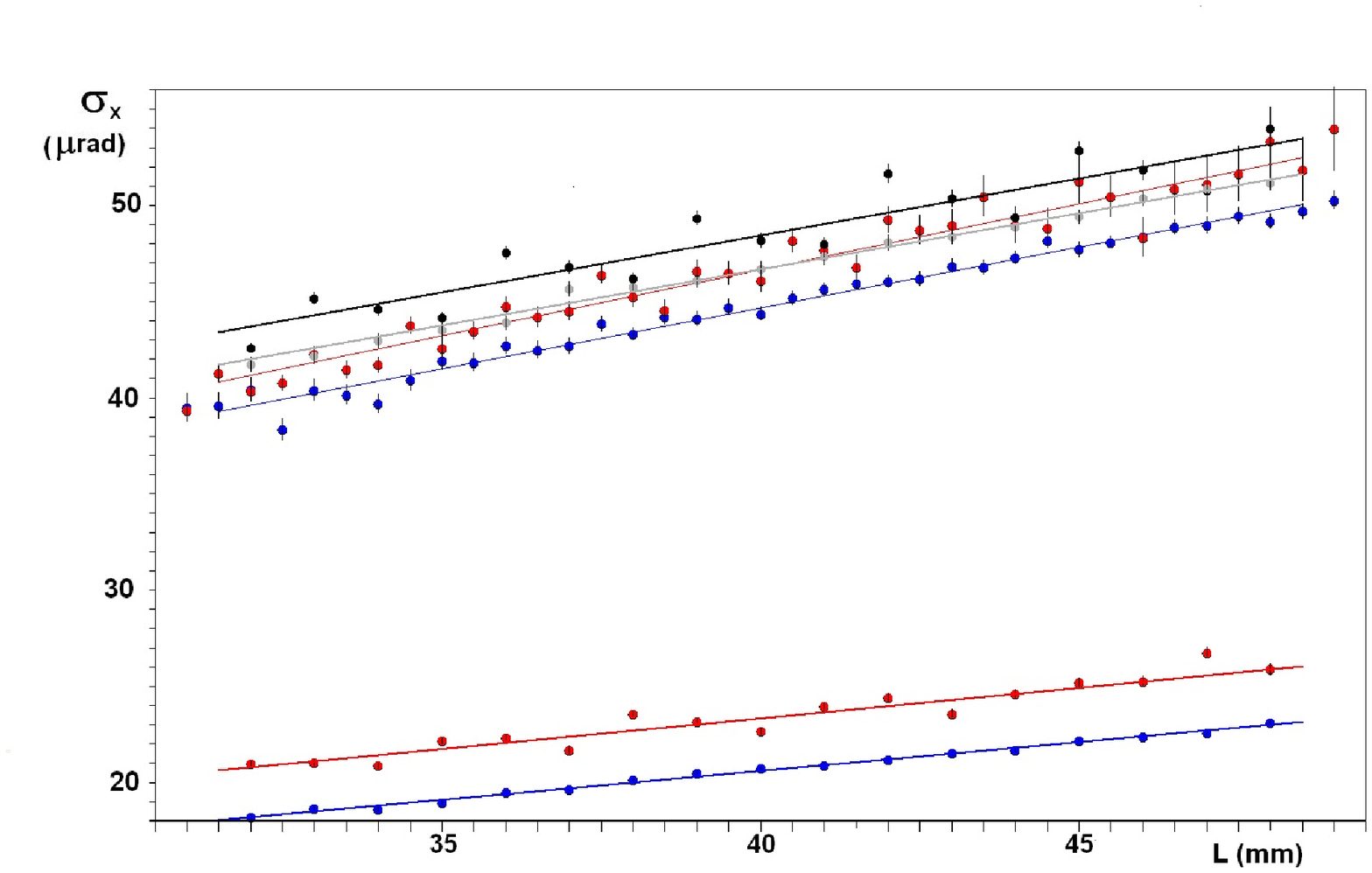}} 
{\caption{
Scattering in the transverse horizontal plane compared with the vertical plane. The upper part of the figure 
shows the rms angles for particles with a momentum of 180 GeV/c. Black and gray dots (and straight lines) 
correspond to the horizontal and vertical planes of crystal 3. Similarly, the red and blue colors for crystal 4.
Below are the data for crystal 1 and particle momentum of 400 GeV/c with the same colors as for crystal 4. 
}}
\end{center}
\end{figure*}
\begin{figure*} 
\begin{center}
\scalebox{0.8}
{\includegraphics{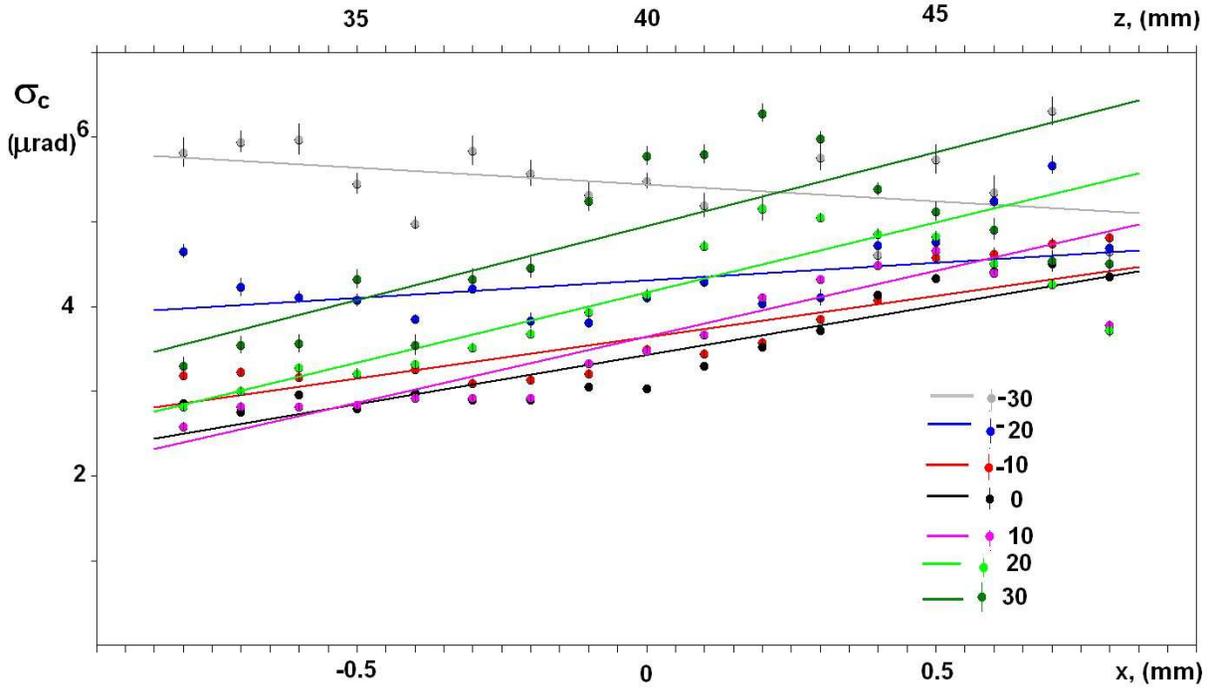}} 
{\caption{
The rms angle of multiple scattering of 400 GeV/c protons  moving in the channeling region
in narrow strips as a function of the crystal length (upper scale)
and the transverse horizontal coordinate $x$. In the lower right corner, the coordinates of the centers of the strips are shown in $\mu$rad. 
}}
\end{center}
\end{figure*}
\begin{figure*} 
\begin{center}
\scalebox{0.8}
{\includegraphics{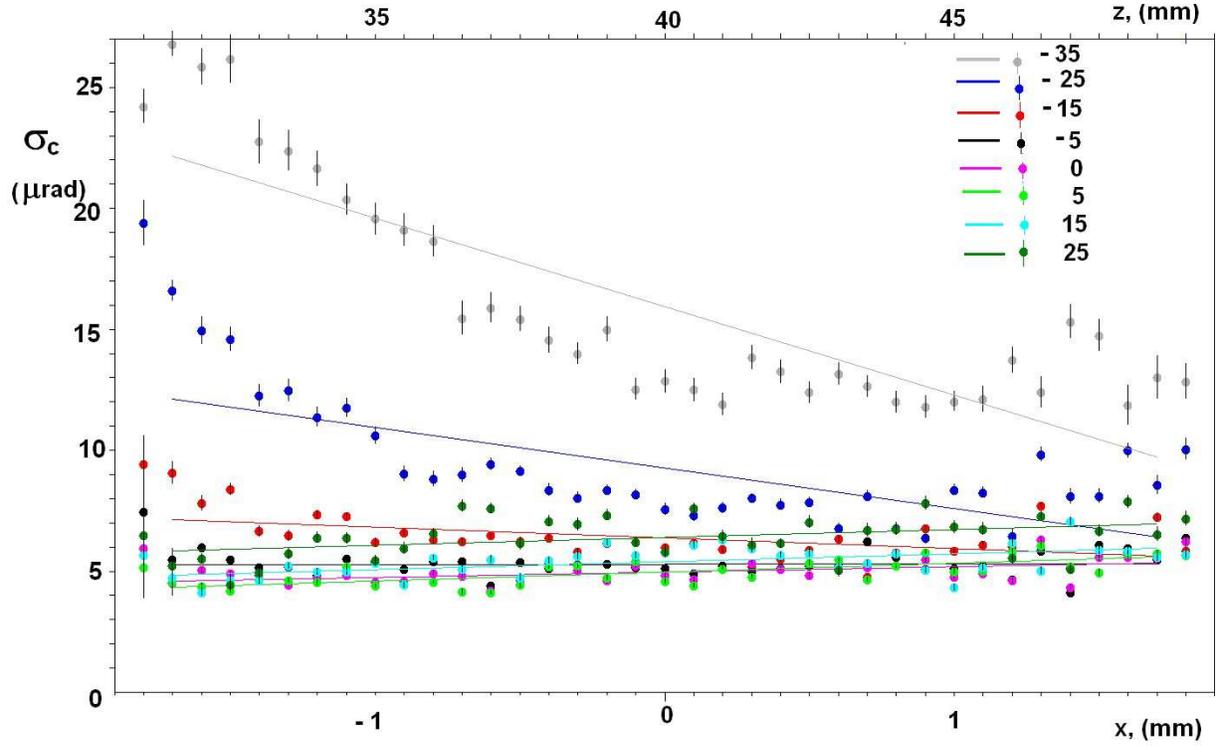}} 
{\caption{
The same as in Fig. 4 but for 180 GeV/c and for crystal 4. 
In the upper right corner, the coordinates of the centers of the strips are shown in $\mu$rad. 
}}
\end{center}
\end{figure*}
\begin{figure*} 
\begin{center}
\scalebox{0.8}
{\includegraphics{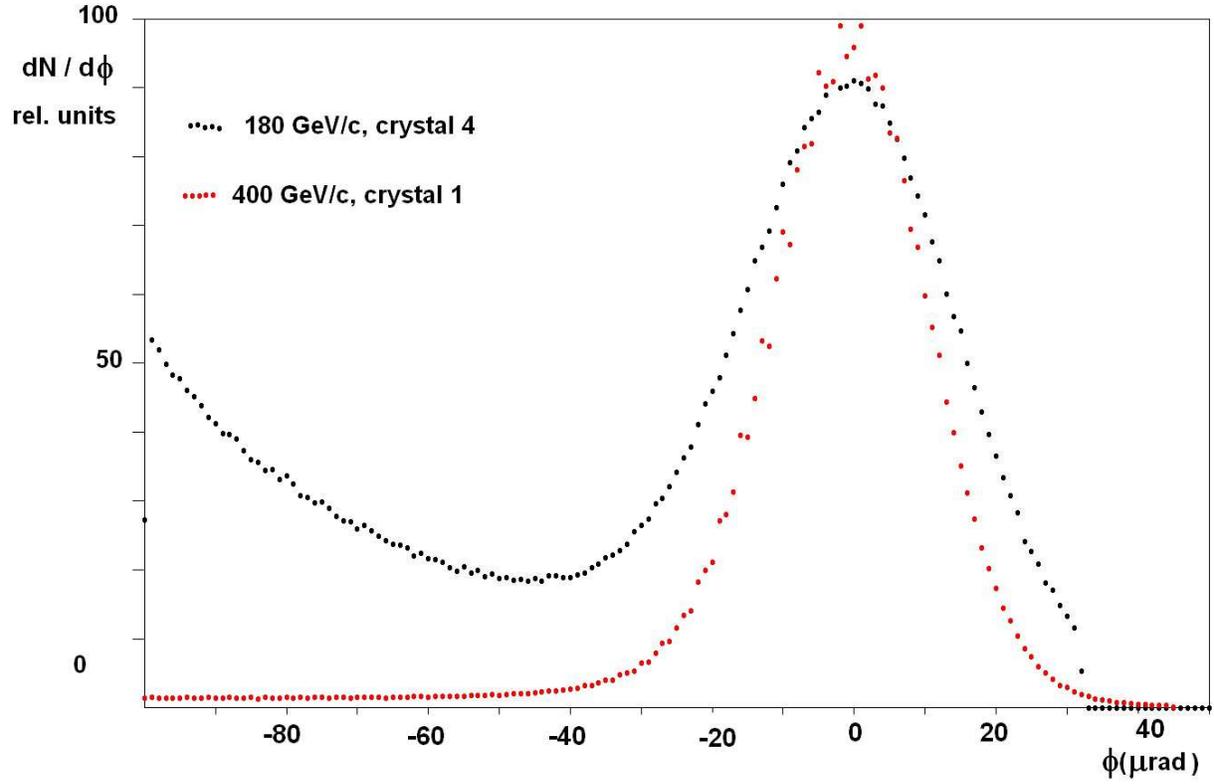}} 
{\caption{
Distribution of particles moving in the channeling region by angle $\phi$ relative to the straight line $ QQ '$ (see Fig. 1b). 
}}
\end{center}
\end{figure*}

\begin{figure*} 
\begin{center}
\scalebox{0.8}
{\includegraphics{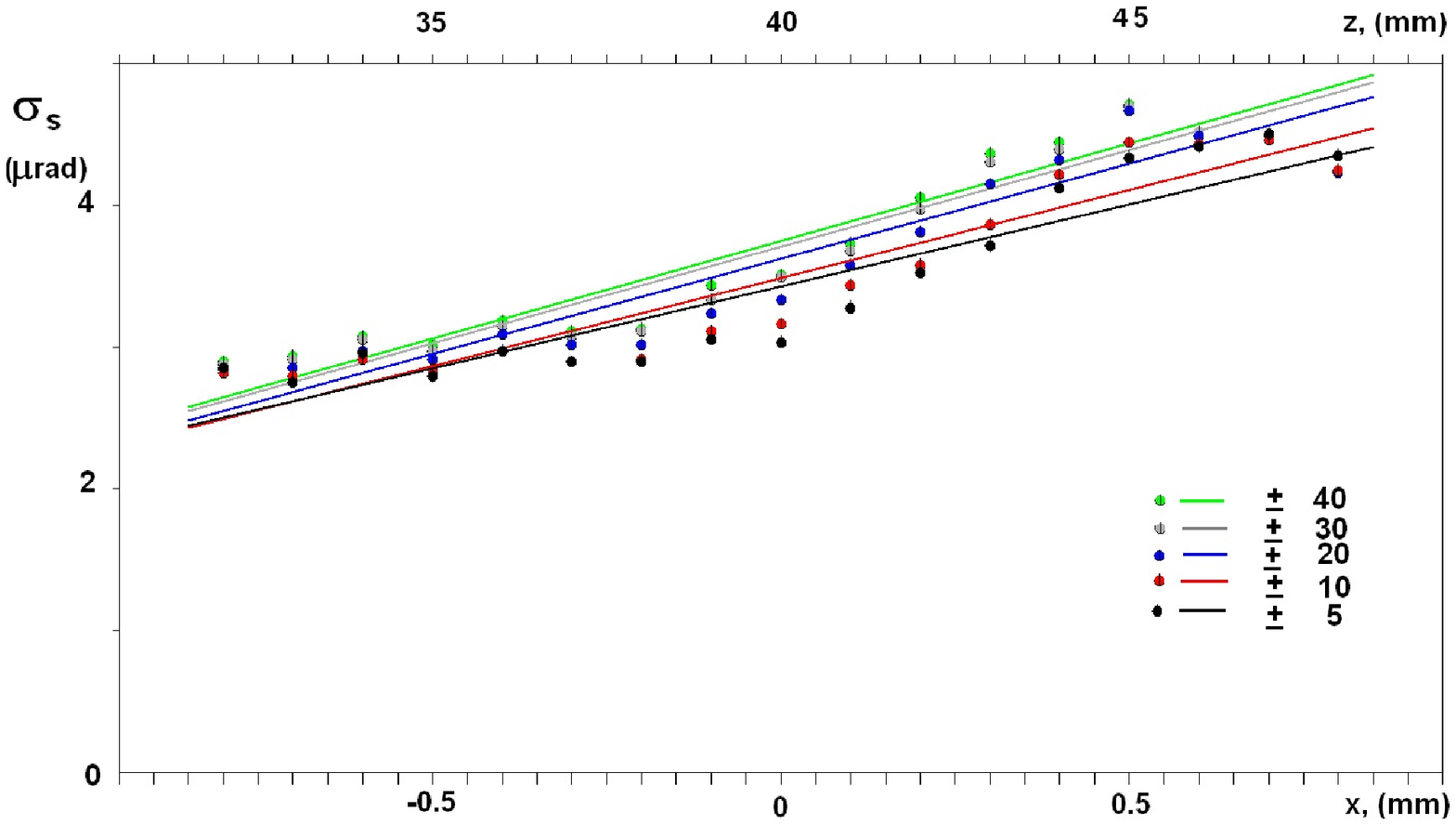}} 
{\caption{
The dependence of the rms scattering angle of 400 GeV/c  protons contained in the range from $- \phi_o $ to $\phi_o$ 
(shown in the figure in $\mu$rad) as a function of the crystal thickness.
}}
\end{center}
\end{figure*}
\begin{figure*} 
\begin{center}
\scalebox{0.8}
{\includegraphics{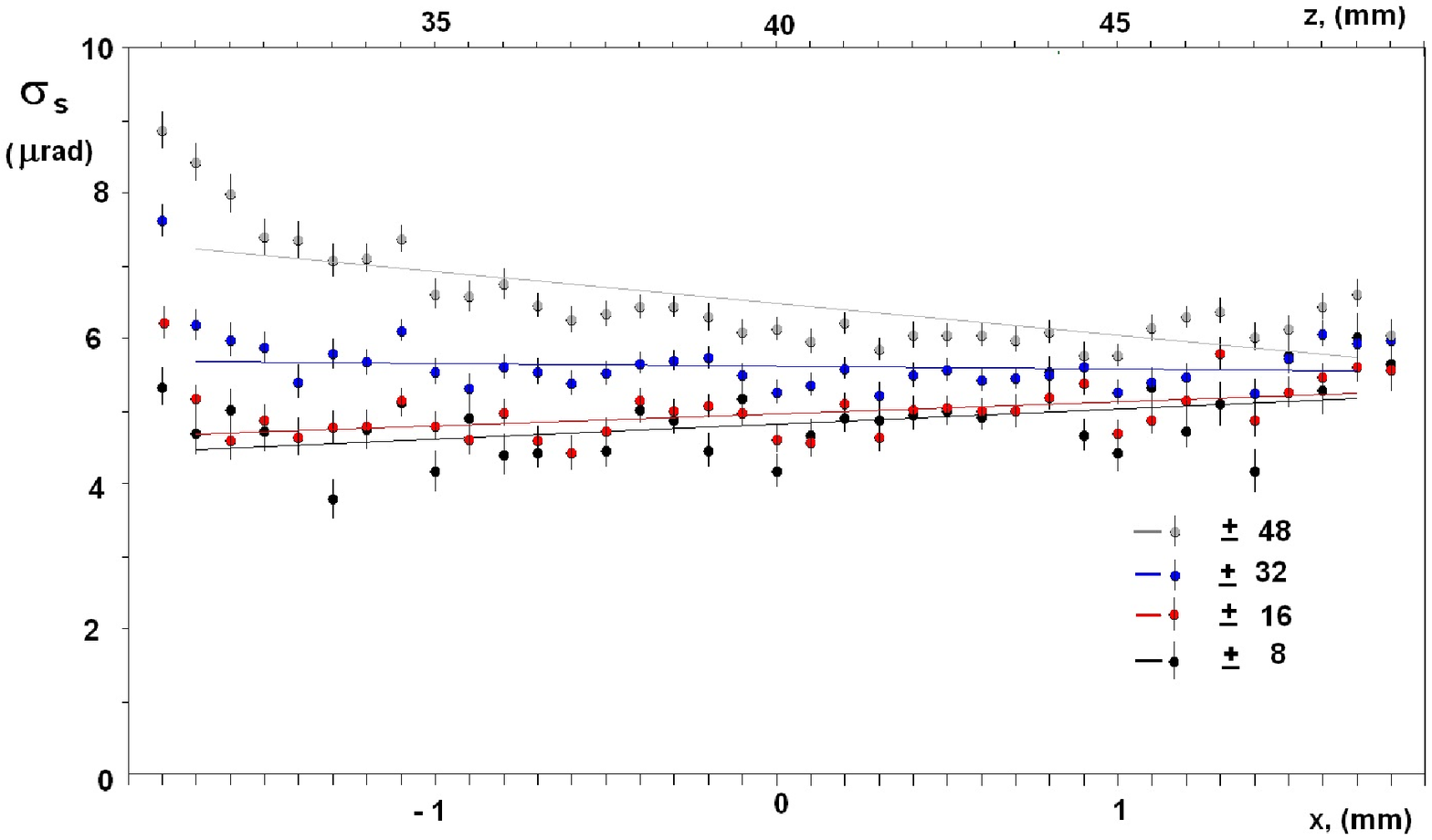}} 
{\caption{
The dependence of the rms scattering angle of 180 GeV/c particles contained in the range from $- \phi_o $ to $\phi_o$ 
(shown in the figure in $\mu$rad) as a function of the crystal thickness.
}}
\end{center}
\end{figure*}
\begin{figure*} 
\begin{center}
\scalebox{0.8}
{\includegraphics{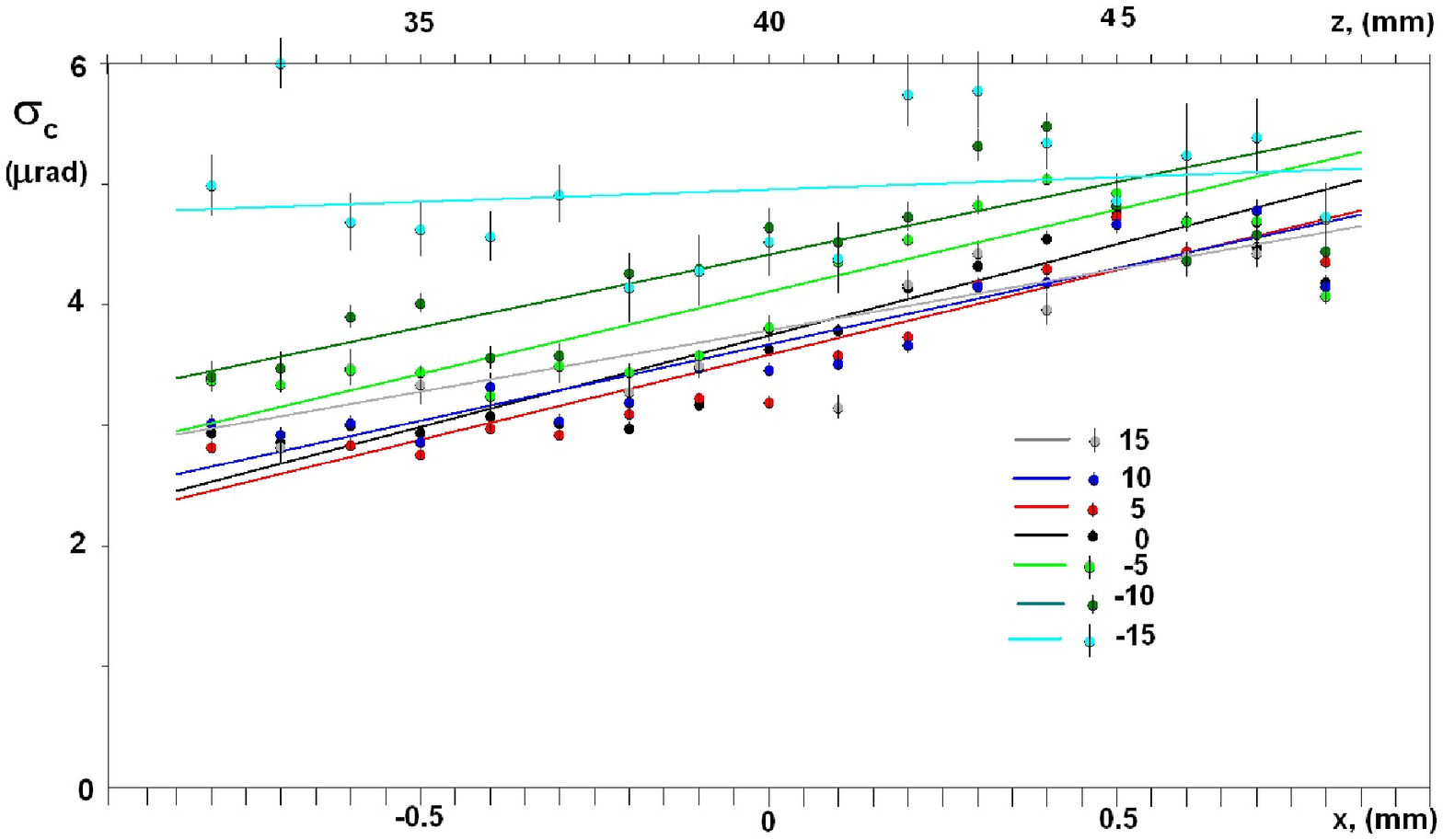}} 
{\caption{
Dependence of the rms scattering angle for a 400 GeV/c proton beam with an initial horizontal entry angle into the crystal $\alpha$ (indicated in the figure in $\mu$rad ) as a function of the thickness. 
}}
\end{center}
\end{figure*}
\begin{figure*} 
\begin{center}
\scalebox{0.8}
{\includegraphics{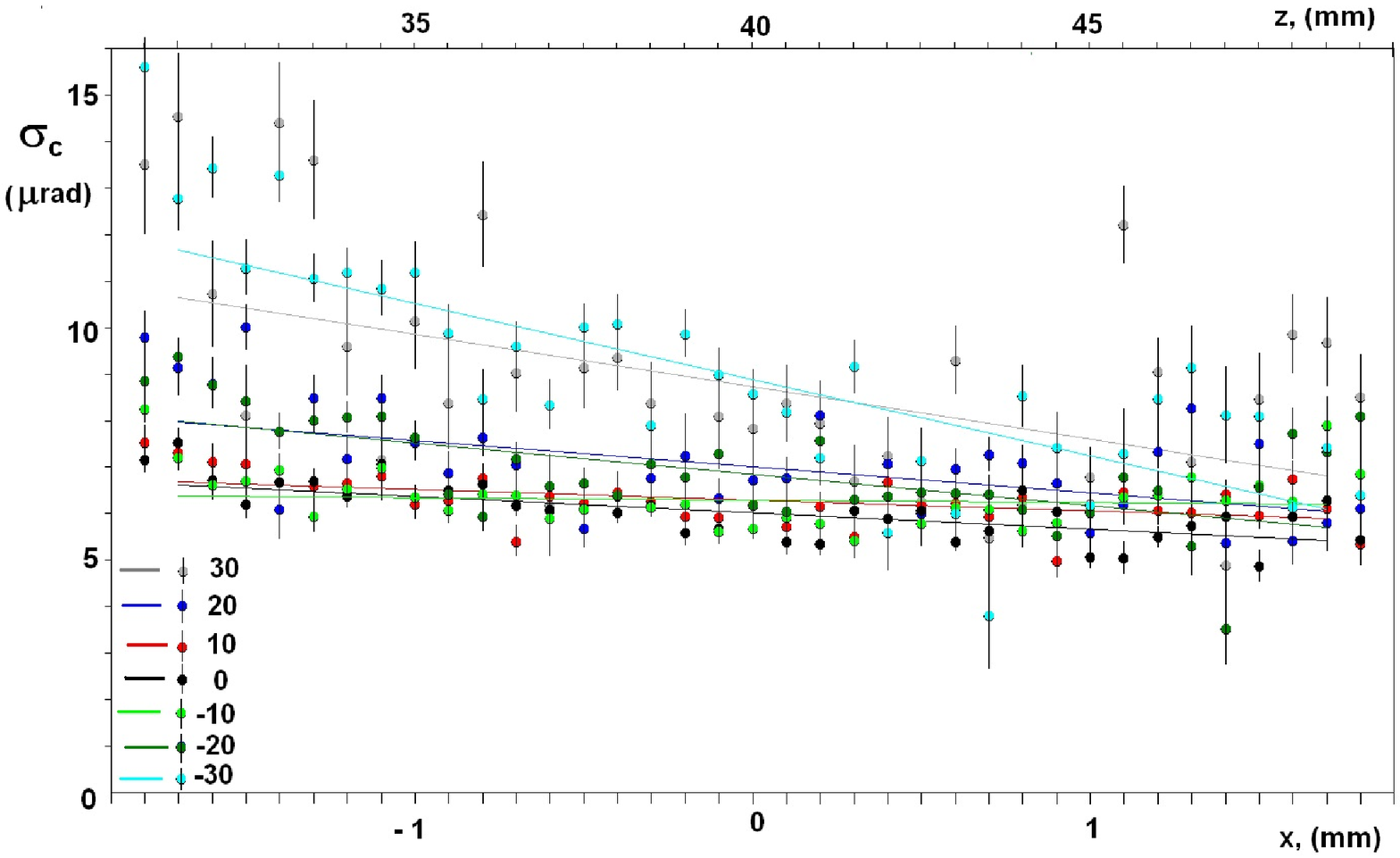}} 
{\caption{
Dependence of the rms scattering angle for a 180 GeV/c  beam with an initial horizontal entry angle  into the crystal 
$\alpha$ (indicated in the figure in $\mu$rad) as a function of the thickness. 
}}
\end{center}
\end{figure*}
\begin{figure*} 
\begin{center}
\scalebox{0.8}
{\includegraphics{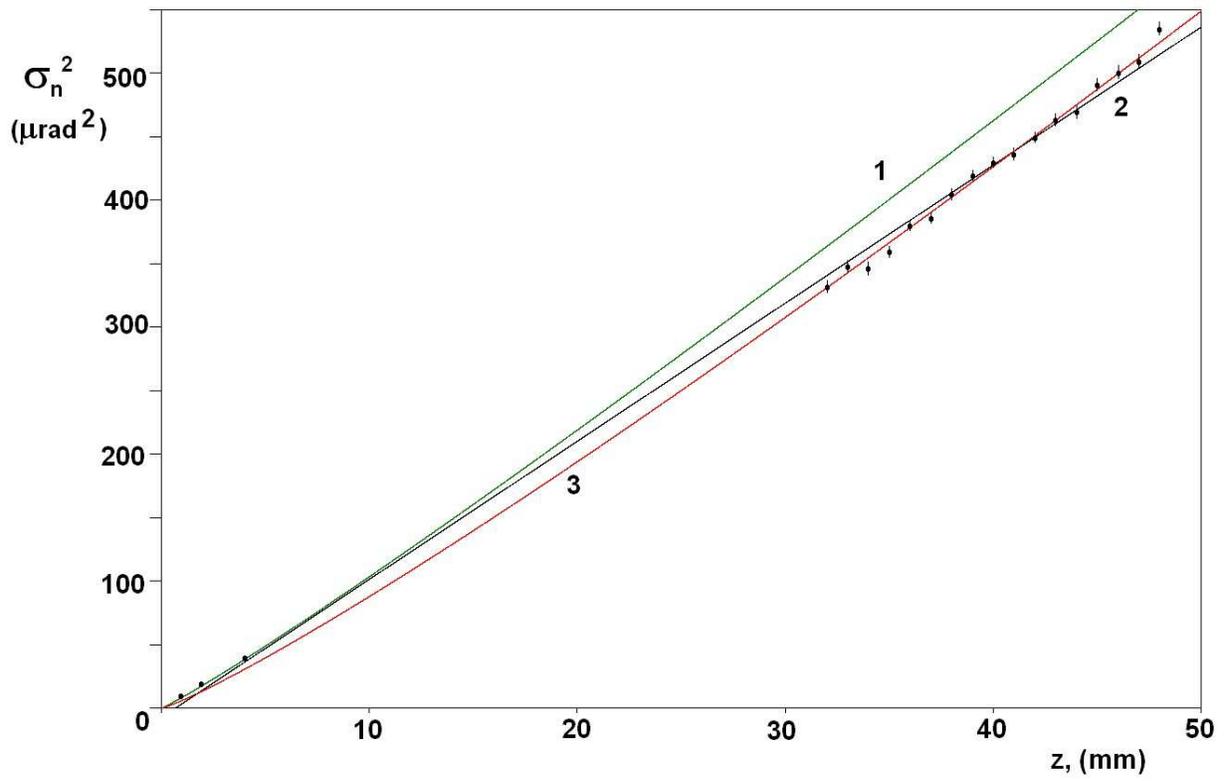}} 
{\caption{
Approximations  of the mean square  multiple scattering angle for 400 GeV protons in  monocrystalline silicon.
The points are the experimental data. Curve 1 corresponds to Eq.(4),  curve 2 is a linear approximation of Eq.(8), 
and curve 3 is Eq.(9)  for $\varepsilon =13.35$ MeV, $\, \omega=0.063$.
}}
\end{center}
\end{figure*}
\begin{figure*} 
\begin{center}
\scalebox{0.8}
{\includegraphics{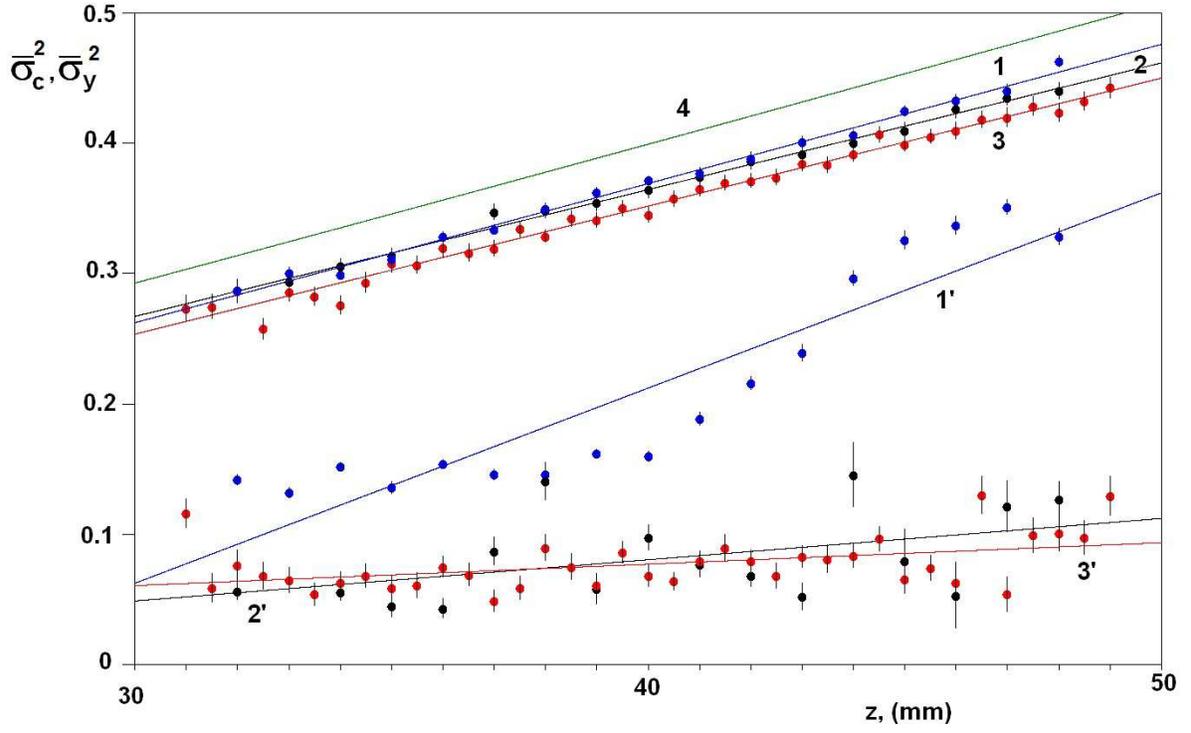}} 
{\caption{
The dependence of the mean square multiple scattering angle
divided by $k^2$ ( $k=13.6$ [MeV] $/ E]$) as a function of thickness.
The curves
1,1' and 2,2' and 3,3'  correspond to  non-channeling and channeling regimes.  The data for channeling regimes  were enlarged by a factor 20. Curve 4 corresponds to Eq.(4).
}}
\end{center}
\end{figure*}

\end{document}